# Finite Temperature QCD on the Lattice*

Kazuyuki Kanaya[a]

[a]Center for Computational Physics and Institute of Physics, University of Tsukuba,
Tsukuba, Ibaraki 305, Japan

Recent progress in lattice QCD at finite temperature is reviewed.

## 1. INTRODUCTION

Clarification of the nature of the finite temperature QCD matter and its temperature driven phase transition or crossover is one of the major goals of QCD calculations on the lattice. For simplicity, I use the term "transition" for both genuine phase transitions and sharp crossovers in this report, unless explicitly specified. In a study of the order of a transition, the scaling analysis is a powerful tool. Unlike for the SU(3) pure gauge theory where the finite spatial size of the lattice is the only cause to smoothen the singularity at the transition point [1], near the chiral transition in QCD the quark mass is expected to play a dominant role provided that the spatial lattice size is not too small [2]. The scaling analysis is particularly important in a study of the chiral transition with two flavors because theoretical studies do not have a definite prediction about the order of the transition in this case [3,4]. This year, new studies along this line were presented.

Most numerical studies in finite temperature QCD have been performed with staggered quarks. Although the flavor structure is much simpler with Wilson quarks, the violation of chiral symmetry due to the Wilson term poses several conceptual and technical difficulties. Recent efforts with Wilson quarks revealed several unexpected properties and a more complicated phase structure than that for staggered quarks [5]. In order to make a reliable prediction for the continuum limit, however, it is essential to remove the dependence on the lattice formulation of quarks in the results. Several fundamental progresses were reported for Wilson quarks at the conference.

In a more realistic case including the strange quark, the status is still not clear enough. With staggered quarks, the Columbia group found a time history that suggests a crossover at a point where $m_s$ is smaller than its physical value and $m_u$ and $m_d$ are larger than their physical values [6]. Because the results for two flavors suggest a crossover at $m_s = \infty$ and at these values of $m_u$ and $m_d$, this implies that the transition in the real world is also a crossover provided that the dependence on the light quark mass is negligible. On the other hand, a recent study of an effective $\sigma$ model suggests that the nature of the transition can sensitively depend on the light quark mass [4,7]. Therefore, it is important to determine the global phase structure in the full coupling parameter space. A new result with Wilson quarks was presented.

For the nature of the finite temperature QCD matter, simulations on larger lattices were made to determine the transition temperature and the equation of state; the behavior of high temperature gluon propagators was studied; quenched calculations on the finite temperature hadronic matter were performed; and properties of a valence quark chiral condensate were studied.

QCD at finite density is another highly important subject especially in phenomenological applications to heavy ion collisions and astrophysical

---

Table 1
Critical exponents of the three dimensional O(4) [10] and O(2) [11] Heisenberg models, the meanfield theory, and $N_F = 2$ QCD with staggered quarks [12].

|  | O(4) | O(2) | MF | Karsch-Laermann |  |
|---|---|---|---|---|---|
| $1/\beta\delta$ | 0.537(7) | 0.602(2) | 2/3 | 0.77(14) | $\beta_c$ |
| $1/\delta$ | 0.2061(9) | 0.2072(3) | 1/3 | $0.21 - 0.26$ | chiral cumulant |
| $1 - 1/\delta$ | 0.7939(9) | 0.7928(3) | 2/3 | 0.79(4) | $\chi_m$ |
| $(1-\beta)/\beta\delta$ | 0.331(7) | 0.394(2) | 1/3 | 0.65(7) | $\chi_t$ |
| $\alpha/\beta\delta$ | $-0.13(3)$ | $-0.003(4)$ | 0 | $-0.07 - +0.34$ | specific heat |

processes. Nevertheless, a reliable numerical simulation is still difficult as a consequence of the complex action problem [8]. A few new studies were reported this year.

In this report I attempt to review these new developments. In Sect. 2, I review recent studies on the chiral transition with two flavors of staggered and Wilson quarks. Sect. 3 deals with the influence of the strange quark. In Sect. 4, we concentrate on other subjects in the QCD thermodynamics. Recent progress at finite chemical potential is discussed in Sect. 5. A brief summary is given in Sect. 6. For the previous status of investigations, I refer the reader to Refs.[9,5,7].

## 2. CHIRAL TRANSITION IN QCD WITH TWO FLAVORS

Understanding the nature of the finite temperature transition in QCD with two degenerate light quarks ($N_F = 2$) is an important step toward the clarification of the transition in the real world. Based on a study of an effective $\sigma$ model Pisarski and Wilczek discussed that the transition in the chiral limit (chiral transition) of QCD with two flavors is either of second order or of first order depending on the strength of the $U_A(1)$ anomaly term at the transition temperature [3]. Three years ago, Wilczek and Rajagopal argued that, in case that the chiral transition is of second order, QCD with two flavors will belong to the same universality class as the three dimensional O(4) Heisenberg model [4]. This provides us with several useful scaling properties that allow us to make clear the order of the transition. Because no known lattice fermions have the full chiral symmetry on finite lattices, the appearance of the O(4) scaling will be a useful touchstone to test the recovery of the full chiral symmetry on the lattice if the chiral transition is of second order.

Scaling properties are described in terms of critical exponents. Consider a spin model at temperature $T$ near the transition temperature $T_c$ and at small external magnetic field $h$. The exponents $\alpha$, $\beta$, $\gamma$, and $\delta$ are defined for the specific heat $C$, the magnetization $M$, and the magnetic susceptibility $\chi$ by

$$C(t, h=0) \sim |t|^{-\alpha} \qquad (1)$$
$$M(t, h=0) \sim |t|^{\beta} \quad (t < 0) \qquad (2)$$
$$\chi(t, h=0) \sim t^{-\gamma} \quad (t > 0) \qquad (3)$$
$$M(t=0, h) \sim h^{1/\delta} \qquad (4)$$

where $t = [T - T_c(h=0)]/T_c(h=0)$ is the reduced temperature. These critical exponents satisfy the following two scaling relations: $\alpha = 2 - \beta(\delta + 1)$ and $\gamma = \beta(\delta - 1)$, so that only two of the four exponents are independent. Some O(4) exponents are listed in Table 1. In QCD, $h$ corresponds to the quark mass and $M$ corresponds to the chiral condensate.

### 2.1. Scaling study with staggered quarks

The conjecture by Wilczek and Rajagopal has been tested first for staggered quarks [12]. Numerical results obtained so far for the transition with $N_F = 2$ staggered quarks are consistent with a second order chiral transition: No signs of discontinuities are found at $m_q a \geq 0.0125$ for $N_t = 4$ and 6, where $N_t$ is the lattice extension in the temporal direction [6,13,14]. Also a finite size scaling for a first order transition is not compatible with the data obtained at $m_q a = 0.01 - 0.0125$ and $N_t = 4$ [14,15].

The last two columns in Table 1 show the nu-merical results for critical exponents by Karsch and Laermann obtained at $m_q a = 0.02 - 0.075$ on an $8^3 \times 4$ lattice [12]. The result for $\delta$ is consistent with the O(4) value, while the results for $\alpha$ and $\beta$ are in disagreement with the O(4) exponents. Clearly, we need data with better accuracy (on larger lattices and at smaller $m_q a$). The Bielefeld group continued their efforts to improve the simulation and preliminary results on larger spatial lattices ($12^3$ and $16^3$) were presented at the conference [16]. The results obtained so far are consistent with those on the $8^3 \times 4$ lattice (cf. Fig. 1). Another test of the O(4) scaling performed by DeTar [9] will be discussed later.

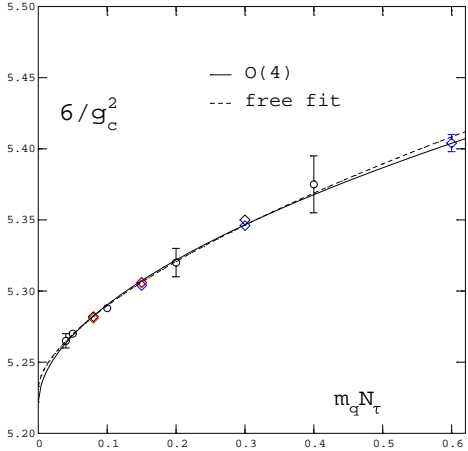

Figure 1. Pseudocritical coupling $\beta_c$ versus $m_q/T$ at $N_t = 4$ [16]. This figure is an update of Fig. 6 in Ref.[12] including new results obtained on $12^3 \times 4$ and $16^3 \times 4$ lattices (a part of diamonds at $m_q/T = 0.08$, 0.13, 0.3 and 0.6).

With staggered quarks, however, several caveats are in order because $N_F = 2$ staggered quarks are realized on the lattice by introducing a fractional exponent to the fermionic determinant of $N_F = 4$ staggered quarks. Therefore, (i) the symmetry in the chiral limit on a lattice with finite lattice spacing is the O(2) symmetry of $N_F = 4$ staggered quarks [17], and not the O(4) symmetry we expect to find in the continuum limit, and (ii) the action is not local. The correct continuum chiral limit with the O(4) symmetry will be obtained only when we first take the continuum limit $\beta \to \infty$ and then take the chiral limit. Choosing a too small $m_q$ compared with the lattice spacing may lead us either to wrong O(2) exponents, or to some non-universal behavior due to the lack of locality.

Another theoretical possibility (both for staggered and Wilson quarks) is the appearance of the meanfield (MF) exponents [18]. Out of the critical region (Ginzburg region), the order parameter may be determined by a MF theory [7]. For example, the critical region in the BCS theory is known to be quite narrow so that many properties of superconductors can be calculated with the MF theory by Ginzburg and Landau. In 3D and 4D Gross-Neveu models, Kocić and Kogut reported the absence of critical regions [18]. Narrowness in the width of the critical region in the BCS theory is closely connected to the fact that Cooper pairs are very loosely bounded. In QCD, because quarks are tightly bounded at low temperatures, we may expect the critical region to be less narrow.

These possible exponents are listed in Table 1. It is important that we are able to make a numerical distinction between these different sets of critical exponents. Unfortunately, O(4) and O(2) exponents are very close to each other.

## 2.2. Phase structure and problems with Wilson quarks

In the case of Wilson quarks, the lattice action lacks chiral symmetry due to the Wilson term. This introduces possible $O(a)$ effects to physical quantities and relations characteristic for chiral symmetry. However, when we are close enough to the continuum limit, we expect that the symmetry breaking effects become sufficiently weak so that we see the O(4) scaling when the chiral transition is of second order. In this case the appearance of the O(4) scaling is a useful test to see if the effects from the Wilson term are small.

Although chiral symmetry is broken explicitly, numerical simulations show that the relation $m_\pi^2 \propto m_q$ at small $m_q$ is well satisfied in the low temperature phase for all values of $\beta$, where $m_q$ is defined via an axial Ward identity [19–21]. It is therefore natural to define the chiral limit $K_c$ by $m_\pi = 0$ or $m_q = 0$ at $T = 0$. The loca-tion of $K_c$ for $N_F = 2$ is shown by the dark-gray line in Fig. 2 collecting recent data from various groups [5]. There exist small discrepancies (at most of the order of 0.01) between the values for $K_c$ from $m_\pi = 0$ and $m_q = 0$ which are attributed to $O(a)$ effects due to the Wilson term, and also to the errors from the extrapolation of $m_\pi^2$ and $m_q$ in $1/K$. The light-gray lines $K_t(N_t)$ indicate the finite temperature transition determined from rapid changes in physical observables. Because the line $K_c$ is the renormalization group flow line for massless QCD, the finite temperature transition for massless QCD is located at the crossing point of $K_t$ and $K_c$. We shall therefore denote this point as the chiral transition point $\beta_{ct}$. Notice that there are possible small ambiguities of $O(a)$ in the location of the $K_c$ line and the point $\beta_{ct}$.

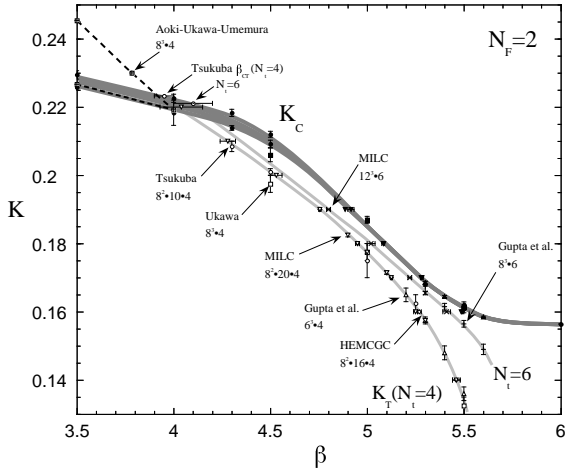

Figure 2. Phase diagram for $N_F = 2$ QCD with Wilson quarks using the standard action. See text for details. Lines are to guide the eyes.

In the high temperature phase, we encounter no singularities even just on the $K_c$ line, where $m_q$ vanishes [21] but the finite temperature pion screening mass $m_\pi(T > 0)$ remains non-zero [22,23]. Also with staggered quarks, we expect a similar absence of singularity on the $m_q = 0$ line in the high temperature phase. In the low temperature phase, on the other hand, there appear small eigenvalues of the quark matrix near $K_c$ which make the number of CG iterations required to invert the quark matrix to a given accuracy, very large [24]. The Tsukuba group performed simulations on the $K_c$ line by reducing $\beta$ from the high temperature side, and looked for a point where singularity first appears in the sense that the number of CG iterations becomes very large. They found that the location of the point is given by $\beta \sim 3.9 - 4.0$ for $N_t = 4$ and $4.0 - 4.2$ for $N_t = 6$. These two points are consistent with the ones obtained from linear extrapolations of the $K_t$ lines [23]. Therefore, they identified these points with $\beta_{ct}(N_t)$.

Recently, Aoki, Ukawa and Umemura stressed that, in understanding the phase structure of Wilson quarks at finite temperatures, it is important to clarify the meaning of the critical line at $T > 0$ [25]. They defined the finite temperature $K_c$ line (which I will denote in the following as the $K_c^{T>0}$ line) by $m_\pi(T > 0) = 0$, and studied it in the parameter space including the region above $K_c$. Let us denote $K_c^{T>0}$ at $T = 0$ as $K_c^{T=0}$. Note that the $K_c^{T=0}$ line defined by $m_\pi(T=0) = 0$ has no width and locates in the $K_c$ band shown in Fig. 2 whose width is cause by $O(a)$ ambiguities mentioned above.

Motivated by their analytic results of the finite temperature 2D Gross-Neveu model, Aoki et al. performed simulations of $N_F = 2$ QCD on an $8^3 \times 4$ lattice and found that the $K_c^{T>0}$ line, which first flows along the $K_c$ line in the low temperature phase, sharply turns back upward in the vicinity of $\beta = 4.0$ (cf. the dashed line in Fig. 2). The study of the Gross-Neveu model suggests that the lower part of the $K_c^{T>0}$ line locates slightly (probably $O(a) \approx O(1/N_t)$) above the $K_c^{T=0}$ line. If this is the case, $m_\pi(T_c > T > 0) \approx O(1/N_t) > 0$ on the $K_c^{T=0}$ line and will vanish only in the continuum limit. The study of the Gross-Neveu model also suggests that, when $N_t$ is increased, the turning point of the $K_c^{T>0}$ line moves along the $K_c$ line toward larger $\beta$ and touch the weak coupling chiral point $K = 1/8$ in the limit $N_t = \infty$. The region surrounded by the $K_c^{T>0}$ line is the parity broken phase proposed by Aoki for $T = 0$ QCD [26]. Concerning the relation between $K_c^{T>0}$ and $K_t$ in QCD, they argued that the $K_t$ line crosses the $K_c^{T=0}$ line and runs past the turning point of the $K_c^{T>0}$ line without touching this line (cf. Fig. 4 in Ref.[25]).

The shape of the $K_c^{T>0}$ line shown in Fig. 2 is in accordance with the absence of singularity in the high temperature phase discussed above. The location of the turning point is consistent with $\beta_{ct}$ estimated by the Tsukuba group. The suggested $O(a)$ gap between the $K_c^{T>0}$ and $K_c^{T=0}$ lines in the low temperature phase was previously observed by the Tsukuba group [23]. We note that the possibility is not excluded that the $K_t$ line touches the turning point of $K_c^{T>0}$. In this case, the crossover $K_t$ becomes a genuine transition at that point.

When we view the lattice QCD at finite $N_t$ and finite $a$ as a statistical system, the phase structure is understood in terms of the critical line $K_c^{T>0}$. On the other hand, because our final goal is to investigate the chiral transition of QCD in the continuum limit, it is important to know the relation between the $K_c$ line and the $K_t$ line and also to trace the recovery of chiral symmetry which is broken on the lattice. This leads us to the $K_c$ line and $\beta_{ct}$ which have $O(a)$ ambiguities as discussed above. In order to recover the finite temperature chiral transition in the continuum limit, the gaps between $K_c^{T>0}$ and $K_c^{T=0}$ and between $K_t$ and the turning point of $K_c^{T>0}$ should be also at most of $O(a)$. The data actually show that these gaps are small to the present numerical accuracy. The continuum limit is not affected by these details.

The important progress made by Aoki et al. is that the existence of the points where $m_\pi(T > 0) = 0$ is explicitly shown in a toy model, and that the phase structure inferred from this model is confirmed in QCD at least partly. This provides us with a more rigid theoretical basis to extract the properties of the chiral transition in the continuum limit from physical observables near $\beta_{ct}$.

From the behavior of physical quantities on the $K_c$ line, the chiral transition at $\beta = \beta_{ct}$ is found to be continuous both for $N_t = 4$ and 6 as in the case of staggered quarks [23]. Nevertheless, it seems less hopeful to observe the desired $O(4)$ scaling on these lattices. The main reason is the fact that the dependence of the transition on $\beta$ and the quark mass are completely different from those we expect when the chiral transition is of second order [22,27,28]: Near the continuum limit, we expect that, as the quark mass increases from the chiral limit, the transition becomes weaker with the quark mass and it becomes strong again when the quark mass is heavy enough to recover the first order transition of the SU(3) gauge theory. Contrary to this expectation, the MILC collaboration found that, when we decrease $K$ from $K_c$ on an $N_t = 4$ lattice, the transition on $K_t$ becomes once very strong at $K \simeq 0.18$ and becomes weaker again at smaller $K$ [22]. On a lattice with $N_t = 6$ they even found a first order transition at $K = 0.17 - 0.19$ [27].

Looking at the phase diagram shown in Fig. 2 more closely, we note that these points of strong transition are just the ones where the $K_t$ lines become once very close to the $K_c$ line due to the sharp bend of the $K_c$ line at $\beta \simeq 5.0$, which is caused by the cross-over phenomenon between weak and strong coupling regions of QCD. Therefore, it seems plausible that the strong transition is a result of lattice artifacts caused by this unusual relation of the $K_t$ and $K_c$ lines [5]. The quark mass $m_q$ also shows an unexpected $N_t$ dependence in the high temperature phase at $\beta \lesssim 5.3$ [22,28], which is due to lattice artifacts, too.

A naive way out of these lattice artifacts is to increase $N_t$ so that we have the chiral transition in the weak coupling region, $\beta_{ct} \gtrsim 5.5$. However, a previous study on an $18^3 \times 24$ lattice suggests that this requires $N_t \gtrsim 18$, although the spatial lattice size is not large enough [23].

## 2.3. Wilson quarks with improved action

Problems discussed in the last part of the previous subsection with Wilson quarks using the standard action have motivated the Tsukuba group to study finite temperature physics using improved actions [28,29]. They studied the renormalization group improved gauge action proposed by Iwasaki [30] and combined it with the standard Wilson quark action. The phase diagram is given in Fig. 3. Unlike for the standard action (cf. Fig. 2), the distance between $K_t$ and $K_c$ grows now monotonically when we increase $\beta$ from the chiral transition point $\beta_{ct} \sim 1.4$. Correspondingly, the $K_t$ transition becomes rapidly weak as we increase $\beta$ starting from $\beta_{ct}$ (cf. Fig. 4). The straight line envelop of $m_\pi^2$ for $N_t = 4$ shown in Fig. 4b coincides with $m_\pi^2$ for $N_t = 8$ and corresponds to the chiral behavior $m_\pi^2 \propto m_q$ in the low temperature phase. Deviation from this line signals the crossover of the system to the high temperature phase. The smoothness of physical observables strongly suggests that the transition is a crossover at $\beta > \beta_{ct}$. We can also see that $m_\pi^2$ on the $K_c$ line monotonically decreases to zero as $\beta \to \beta_{ct} + 0$ [29], implying that the chiral transition is continuous. It has also been shown that the lattice artifact in $m_q$ in the high temperature phase mentioned before disappears with this improved action.

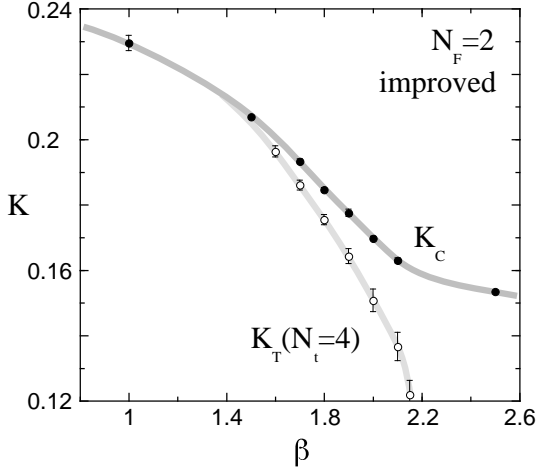

Figure 3. Phase diagram for $N_F = 2$ QCD with Wilson quarks coupled to a RG improved gauge action [29]. $K_c$ is the chiral limit determined by $m_\pi^2 = 0$ on an $8^4$ lattice and $K_t$ is the finite temperature transition line obtained on an $8^3 \times 4$ lattice. Lines are to guide the eyes.

These nice properties which are in accordance with naive expectations encouraged the Tsukuba group to begin a scaling study with Wilson quarks. A key quantity to study scaling properties is the chiral condensate (magnetization). The naive definition of $\langle \bar{\Psi}\Psi \rangle$ is not so useful because the chiral symmetry is explicitly broken due to the presence of the Wilson term. A proper subtraction and a renormalization are required to obtain the correct continuum limit. Just like $m_q$ [19,20], a properly subtracted $\langle \bar{\Psi}\Psi \rangle$ can be defined via an axial Ward identity [19]:

$$\langle \bar{\Psi}\Psi \rangle_{\text{sub}} = 2m_q a Z \sum_x \langle \pi(x)\pi(0) \rangle \quad (5)$$

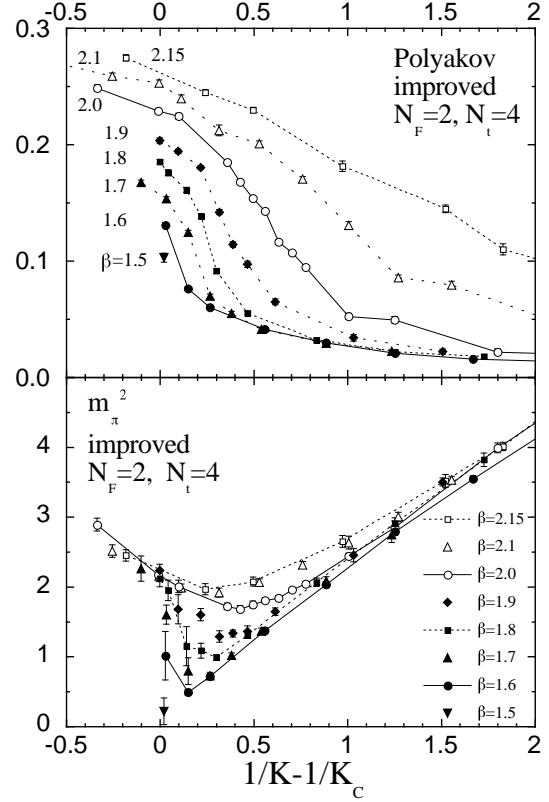

Figure 4. The Polyakov loop and the pion screening mass with a RG improved action at various $\beta$ on an $8^3 \times 4$ lattice [29].

where $Z$ is the renormalization coefficient. This definition is consistent with the identification of the magnetization by Rajagopal and Wilczek [4]. $\langle \bar{\Psi}\Psi \rangle_{\text{sub}}$ was shown to have a non-vanishing value in the chiral limit in the confining phase of quenched QCD [31] and QED [32]. It is easy to see that Eq.(5) holds also with improved gauge actions by replacing $Z$. For our purpose, it is enough to use the tree value: $Z = (2K)^2$.

We now study $\langle \bar{\Psi}\Psi \rangle_{\text{sub}}$ at finite temperature. Results for $N_t = 4$ with the improved action are shown in Fig. 5. From the universality argument we expect that magnetization $M$ near the transition point can be described by a single scaling

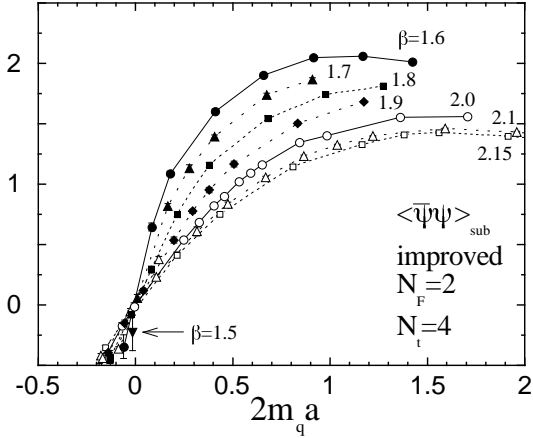

Figure 5. Subtracted chiral condensate $\langle\bar{\Psi}\Psi\rangle_{\text{sub}}$ as a function of $2m_q a$ for Wilson quarks with a RG improved action on an $8^3 \times 4$ lattice [29].

function:

$$M = h^{1/\delta} f(t/h^{1/\beta\delta}) \qquad (6)$$

where $h$ is the external magnetic field ($\propto m_q$) and $t$ is the reduced temperature. DeTar tested this scaling for the case of two flavor staggered quarks [9] and found that data are consistent with O(4) and O(2) scaling.

Fig. 6 shows the result for the scaling function $f(t/h^{1/\beta\delta}) = \langle\bar{\Psi}\Psi\rangle_{\text{sub}}/h^{1/\delta}$ with the identification $M = \langle\bar{\Psi}\Psi\rangle_{\text{sub}}$, $h = 2m_q a$ and $t = \beta - \beta_{ct}$. With fixing the exponents to O(4) and MF values given in Table 1, we adjust $\beta_{ct}$ to obtain the best fit. With the O(4) exponents, we find that the scaling ansatz works remarkably well as shown in Fig. 6a. The resulting $\beta_{ct} = 1.34(3)$ is slightly smaller than $\simeq 1.4$ which is the value obtained by a linear extrapolation of the $K_t$ line (cf. Fig. 3) and of $m_\pi^2$ curves on $K_c$ and $K_t$ [29]. However, the O(4) universality predicts [4] that $\beta_c(m_q) - \beta_{ct} \propto m_q^{1/\beta\delta}$ and $m_\pi^2 \sim (\beta_c - \beta_{ct})^\gamma$ on $K_c$ and $K_t$ with $\gamma \simeq 1.4$, i.e. these lines should bend slightly near the chiral transition point to give a smaller $\beta_{ct}$. Therefore, we conclude that $\beta_{ct} = 1.34(3)$ is consistent with the data. On the other hand, fixing the exponents to the MF values makes the fit less beautiful. The best fit shown in Fig. 6b is obtained with $\beta_{ct} = 1.48(3)$. Contrary to the case of the O(4) exponents, $\beta_{ct} = 1.48(3)$

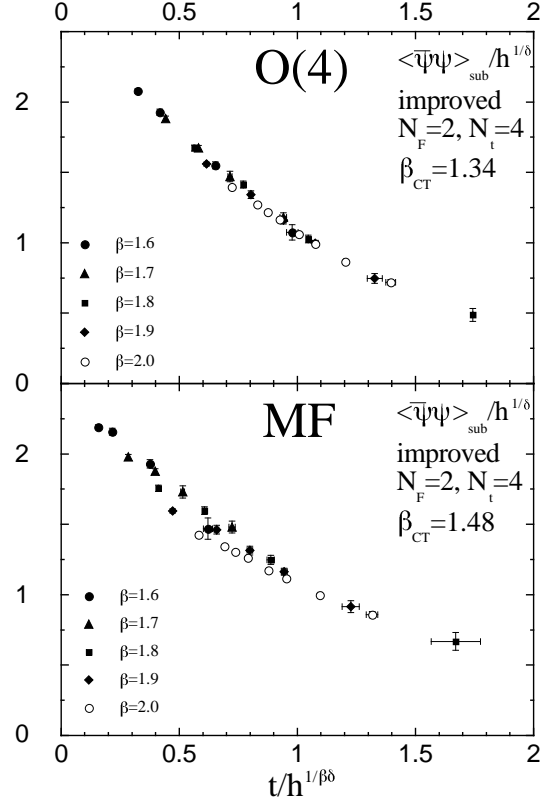

Figure 6. Best fits for the scaling function with O(4) and MF exponents. The plot contains all data of Fig. 5 within the range $0 < 2m_q a < 0.9$ and $\beta \leq 2.0$.

is too large to be accepted.

In conclusion, the data obtained with the improved action are consistent with the predicted O(4) scaling but inconsistent with the MF scaling. The success of this scaling test strongly suggests that the chiral transition is of second order in the continuum limit. It also indicates that, with the improved action, the chiral violation due to the Wilson term is sufficiently weaker than that introduced by the non-vanishing $m_q$ at least for $m_q$'s studied here. For more confirmation of these results, direct extraction of each critical exponent is required.

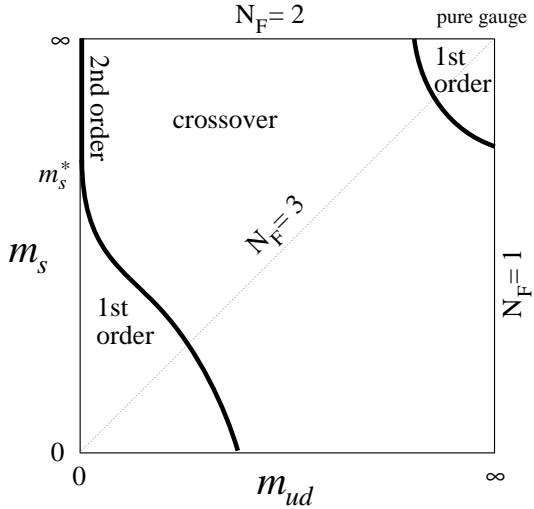

Figure 7. Map of expected nature of the QCD transition for $N_F = 2 + 1$ QCD as a function of the u and d quark mass $m_{ud}$ and the s quark mass $m_s$.

## 3. INFLUENCE OF THE STRANGE QUARK

For QCD with $N_F \geq 3$, Pisarski and Wilczek predicted a first order chiral transition [3]. This was confirmed by numerical simulations both with staggered [6,33] and with Wilson quarks [23]. Off the chiral limit, the first order transition smoothens into a crossover at sufficiently large $m_q$. Therefore, the nature of the transition sensitively depends on $N_F$ and $m_q$. This means that, in order to study the nature of the transition in the real world, we should include the s quark properly whose mass $m_s$ is of the same order of magnitude as the transition temperature $T_c \simeq 100 - 200$ MeV.

### 3.1. Expected phase structure

Following Brown et al. [6], we summarize in Fig. 7 what we expect about the nature of the finite temperature transition as a function of quark masses neglecting the mass difference among u and d quarks ($N_F = 2 + 1$). When all quarks are heavy, the transition is of first order as observed in the SU(3) pure gauge theory. The limit $m_s = \infty$ corresponds to the case $N_F = 2$ discussed in Sect. 2 where we found strong evidence for second order transition at $m_{ud} = 0$. For $m_{ud} = m_s$ ($N_F = 3$), the transition is of first order in the chiral limit. Therefore, on the axis $m_{ud} = 0$, we have a tricritical point $m_s^*$ where the second order transition at large $m_s$ turns into first order [4]. For $m_s > m_s^*$, the second order transition line follows the $m_{ud} = 0$ axis and, for $m_s < m_s^*$, is suggested to deviate from the vertical axis according to $m_{ud} \propto (m_s^* - m_s)^{5/2}$ [7]. Also predicted in Ref.[4] is the MF scaling (with logarithmic corrections) near the tricritical point.

Our main goal of investigations with the s quark is to determine the position of the physical point in this map. I discuss the results of the simulations obtained so far in the next subsection.

### 3.2. Simulations

Using staggered quarks, the Columbia group studied this issue five years ago on a $16^3 \times 4$ lattice [6] extending early studies on smaller lattices [33,34]. For the degenerate $N_F = 3$ case, $m_{ud} = m_s \equiv m_q$, they found a first order signal for $m_q a = 0.025$ at $\beta = 5.132$. For $N_F = 2 + 1$, they obtained a time history suggesting a crossover for $m_{ud}a = 0.025$, $m_s a = 0.1$ at $\beta = 5.171$. Their study of hadron spectrum at this point on a $16^3 \times 24$ lattice gives $m_K/m_\rho$ smaller than the experimental value suggesting that this $m_s$ is smaller than its physical value. At the same time, their large $m_\pi/m_\rho$ suggests that their $m_{ud}$ is larger than the physical value. This implies that the physical point is located in the crossover region unless the second order transition line, which has a sharp $m_{ud}$ dependence near $m_s^*$ (cf. Fig. 7), crosses between the physical point and the simulation point.

Recently, the Tsukuba group studied this issue with Wilson quarks on $8^2 \times 10 \times 4$ and $12^3 \times 4$ lattices [35]. For $N_F = 3$, they performed simulations along the $K_t$ line shown in Fig. 8 and found that the first order signal observed in the chiral limit [23] persists for $\beta \leq 4.7$ (cf. Fig. 9), while no clear two state signals are observed at $\beta = 5.0$ and 5.5. Using $a^{-1} \sim 0.8$ GeV for $\beta \lesssim 4.7$ and $a^{-1} \sim 1.0$ GeV for $\beta = 5.0$, as obtained from the rho meson mass extrapolated to $K_c$, these two cases correspond to $m_q \lesssim 140$ MeV and $m_q \gtrsim 250$

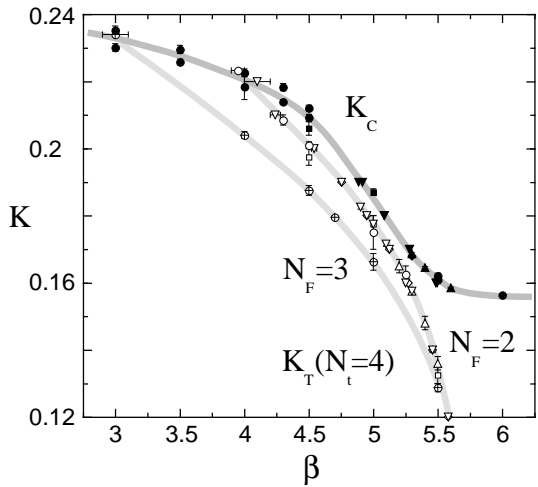

Figure 8. Phase diagram for $N_F = 3$ with Wilson quarks using the standard action [35]. For comparison, the $K_t(N_t = 4)$ line for $N_F = 2$ is included.

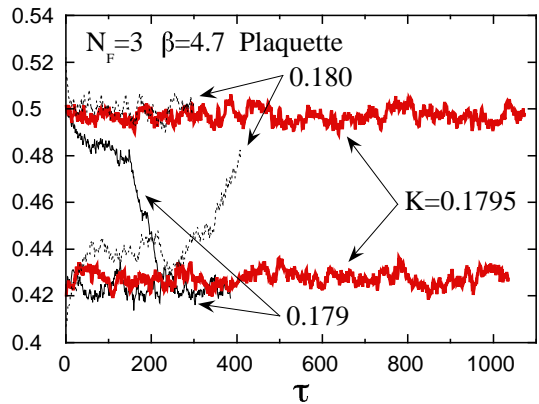

Figure 9. Time history of the plaquette for $N_F = 3$ at $\beta = 4.7$ on a $12^3 \times 4$ lattice with Wilson quarks using the standard action [35].

MeV, respectively. (The physical s quark mass, giving $m_\phi = 1.02$ GeV, was shown to be about 150 MeV with their normalization of $m_q$.) For $N_F = 2 + 1$, first order signals are observed for $m_{ud} \sim 0$ ($K_{ud} = K_c$) at both $m_s \sim 150$ and 400 MeV (corresponding values for $K_s$ are determined using the $a^{-1}$ values mentioned above and the $m_q a$ data for $N_F = 2 - 3$). A recent study of hadron spectroscopy by the Tsukuba group for $N_F = 2 + 1$ QCD at $\beta = 3.5$ on an $8^3 \times 10$ lattice shows that $m_\phi \sim 1.03(5)$ GeV at the simulation point with $m_s \sim 150$ MeV, verifying that this simulation point is very close to the physical point in this sense. Their results on the nature of the transition are summarized in Fig. 10. The physical point clearly falls in the first order region.

Although both staggered and Wilson simulations give a phase structure qualitatively consistent with Fig. 7, Wilson quarks tend to give larger values for critical quark masses (measured by $m_\phi/m_\rho$ etc.) than those with staggered quarks. This leads to the difference in the conclusions about the location of the physical point. However, because both of these studies discuss that the deviation from the continuum limit is large at $N_t = 4$, we should certainly make a calculation at larger $N_t$ [36] or with an improved action in order to draw a definite conclusion about the nature of the QCD transition in the real world. At present, increasing $N_t$ is quite painful especially with Wilson quarks because of the reason discussed in Sect. 2.2. Therefore, the Tsukuba group applied their improved action to study these issues. Some preliminary results were reported at the conference [29], so far with no conclusion about values for critical quark masses.

## 4. QCD THERMODYNAMICS

### 4.1. Transition temperature

The value of the transition temperature $T_c$ in physical units is of great importance for phenomenological applications and for scaling tests.

For the case of the SU(3) gauge theory, Boyd et al. presented new data for $\beta_c$ obtained on $32^3 \times 8$ and $32^3 \times 12$ lattices: $\beta_c = 6.0609(9)$ and $6.3331(13)$, respectively [37]. These values are significantly larger than those obtained previously on spatially small ($N_s/N_t \lesssim 2$) lattices [38] and remove most of the claimed discrepancy [39] between $T_c/\sqrt{\sigma}$ for $N_t \geq 8$ on spatially small lattices and that for $N_t = 4$ and 6 using data for $\beta_c$ obtained on big spatial lattices [40,41] (cf. Fig. 11). Now we can extrapolate toward the continuum limit much more safely: Using the infinite

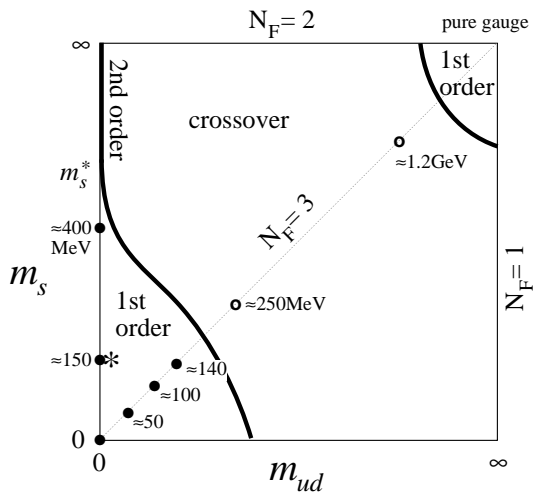

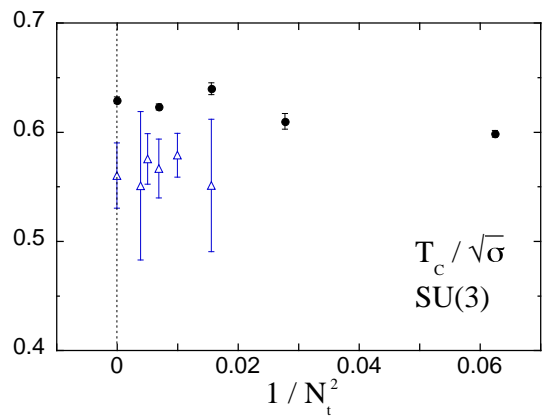

Figure 10. Same as Fig. 7 with the results of simulations with Wilson quarks using the standard action [35]. First order signals are observed at the points marked by filled circles, while no clear two state signals are found at the points represented by the open circles. The values of quark mass in physical units are computed using $a^{-1} \sim 0.8$ GeV for $\beta \leq 4.7$ and $a^{-1} \sim 1.0(1.8)$ GeV for $\beta = 5.0(5.5)$ determined by $m_\rho(T=0, K_c) = 770$ MeV. The real world corresponds to the point marked by the star.

volume $\beta_c$ values for $N_t = 4 - 12$ [40,41,37], Boyd et al. estimated

$$T_c/\sqrt{\sigma} = 0.629(3), \tag{7}$$

for $N_t = \infty$, which is much larger than the previous estimate 0.56(3) using data for $N_t \geq 8$ on small spatial lattices [39]. With $\sqrt{\sigma} = 420$ MeV, this gives $T_c \simeq 260$ MeV.

For the case with dynamical quarks, the spatial lattice size is still small ($N_s/N_t \simeq 2$) for $N_t \geq 6$. This may cause a slight underestimation of $T_c$. The previous status for the transition temperature is summarized in Ref.[9]. At the conference, Blum reported the results of new simulations with $N_F = 2$ staggered quarks performed by the MILC collaboration on $12^3 \times 6$ and $12^4$ lattices [42]: They found that $T_c \simeq 140 - 160$ MeV$_\rho$ in the studied range of $m_q a$ ($m_\pi^2/m_\rho^2 \simeq 0.1 - 0.2$) both for $N_t = 4$ and 6 as can be inferred from the energy density and the pressure plotted in Fig. 12

Figure 11. Transition temperature in the SU(3) gauge theory in units of the square root of the string tension. Filled circles are for the infinite spatial volume limit using data for $\beta_c$ obtained on big spatial lattices [40,41,37], and triangles are for previous estimates using data on small lattices compiled in Ref.[39]. The points at $1/N_t^2 = 0$ are the results of extrapolations [37,39].

(the results for the equation of state will be discussed in the next subsection). Here, "MeV$_\rho$" [9] indicates that $a^{-1}$ is determined by setting $m_\rho = 770$ MeV at $T = 0$ and at each simulation point, i.e. not in the chiral limit [43]. DeTar reported preliminary results of the MILC collaboration from a $24^3 \times 12$ lattice [44]: $T_c \simeq 143 - 154$ MeV$_\rho$ for $m_q a = 0.008$ ($m_\pi^2/m_\rho^2 = 0.2 - 0.3$) [44]. These results are consistent with the previous observation that $T_c$ in MeV$_\rho$ for $N_F = 2$ staggered quarks is stable, to the accuracy of about 20%, for a wide range of $m_\pi^2/m_\rho^2$ and $N_t$ [9]. For QCD with Wilson quarks, $N_t \geq 6$ is required to get a stable $T_c$ [9]. No new results for $N_t \geq 6$ were reported this year.

### 4.2. Equation of state

Energy density $\epsilon$ and pressure $p$ are defined by

$$\epsilon = -\frac{1}{V}\left(\frac{\partial}{\partial(1/T)} \ln Z\right)_V \tag{8}$$

$$p = T\left(\frac{\partial}{\partial V} \ln Z\right)_T \tag{9}$$

where $Z$ is the finite temperature partition function of the theory and $V$ is the spatial volume in

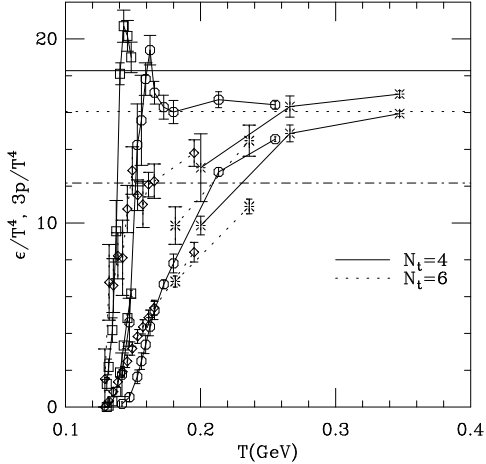

Figure 12. Comparison of the equation of state for $N_F = 2$ at $N_t = 4$ (solid lines) and 6 (dashed lines) [42]. Upper lines are for $\epsilon/T^4$ and lower lines for $3p/T^4$. The results shown are for $m_q a = 0.0125$ (diamonds), 0.025 (octagons), and 0.1 (squares). Bursts are extrapolations to $m_q = 0$. The horizontal lines give the Stephan-Boltzmann law for $N_t = 4$, 6, and the continuum (lowest line). Scale for temperature $T$ was fixed by $m_\rho(T=0) = 770$ MeV at each simulation point.

physical units. In numerical studies on the lattice, it is useful to consider the combination

$$\begin{aligned}\epsilon - 3p &= -\frac{T}{V}[(\frac{1}{T}\frac{\partial}{\partial(1/T)} + 3V\frac{\partial}{\partial V})\ln Z] \\ &= -\frac{T}{V}a\frac{\partial}{\partial a}\ln Z \\ &= -\frac{T}{V}\sum_i (a\frac{\partial g_i}{\partial a})(\frac{\partial}{\partial g_i}\ln Z),\end{aligned} \quad (10)$$

where $g_i$'s are the coupling parameters of the theory [45,41]. In Eq.(10), $\partial \ln Z/\partial g_i$ can be computed numerically as an expectation value of operators on the lattice. $\epsilon - 3p$ is called "the interaction measure" because it vanishes in free Stefan-Boltzmann (SB) gases (see, however, the discussion below). On the other hand, $p$ can be evaluated on the lattice using the following integral representation derived from a thermodynamic relation, $p = -f$, with $f = -T\ln Z/V = -T\partial\ln Z/\partial V$ the free energy density for homogeneous systems [45]:

$$pa^4 = -fa^4 = p(g_i^0)\,a^4 - \int_{g_i^0}^{g_i} dg_i \frac{\partial(fa^4)}{\partial g_i}. \quad (11)$$

Usually, the $T = 0$ contribution is subtracted and $g_i^0$ is chosen so that the system is in the low temperature phase, to obtain $p(g_i^0) \simeq 0$. A convenient choice for the coupling parameter $g_i$ is $\beta$, while with dynamical quarks $m_q a$ is reported to be useful [43].

For the SU(3) gauge theory, Boyd et al. studied $\epsilon$ and $p$ on $N_t = 4 - 8$ lattices with large spatial volumes ($N_s/N_t \geq 4$) [37]. For the nonperturbative beta-function required in Eq.(10) and also in fixing the temperature scale, they used data from a MCRG study by the QCDTARO collaboration [46] and the data for $\beta_c$ discussed in Sect. 4.1. They extrapolated the results for $p$ etc. to the continuum limit $N_t = \infty$ using the volume dependence of a SB gas to leading order in $1/N_t^2$. Their results show that the data for $N_t = 6$ and 8 are already close to their $N_t = \infty$ limit, while the discrepancy is significant for $N_t = 4$ data (about 20% larger in $p/T^4$ at $T \simeq 2T_c$). Nevertheless, the previously reported sizable deviation of $p/T^4$ from the SB limit for $N_t = 4$ remains also in the $N_t = \infty$ limit at $T \lesssim 4T_c$ (cf. Fig. 13).

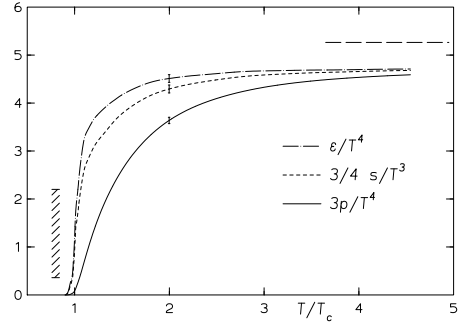

Figure 13. Extrapolation to the continuum limit for the energy density, the entropy density and the pressure using data for the SU(3) gauge theory at $N_t = 6$ and 8 [37]. The dashed horizontal line shows the ideal gas limit. Separation of the data at $T_c$ to high and low temperature phases was not attempted. The hatched vertical band indicates the size of the discontinuity in $\epsilon/T^4$ [41].

For the case of $N_F = 2$ staggered quarks, Blum reported results for $\epsilon$ and $p$ by the MILC collaboration on a $12^3 \times 6$ lattice [42] using the nonperturbative beta-functions for $\beta$ and $m_q a$ studied in Ref.[43]. At $\beta = 5.45$ and 5.53, they performed an extrapolation of $\epsilon$ and $p$ to $m_q = 0$. Their results, together with those for $N_t = 4$, are shown in Fig. 12. Similar to the case of the pure gauge theory, we see that the $p$ and $\epsilon$ values for $N_t = 6$ are by $15 - 20\%$ smaller than those for $N_t = 4$ at $T \sim 1.5 T_c$. The approach to the SB limit is slow on both lattices.

This slow development of $p/T^4$ above $T_c$ as well as the broad tail of the interaction measure at high temperatures has been considered as a sign for large nonperturbative interactions in the quark gluon plasma.

Quite recently, Asakawa and Hatsuda [47] pointed out that the major part of these $T$ dependences in the high temperature phase are caused by the conventional normalization $p(\text{low } T) = 0$, and the slow approach of $p(T)/T^4$ to $p_{\text{SB}}(T)/T^4$ by itself does not necessarily imply large nonperturbative effects at those temperatures: Note that we have a freedom to shift $\epsilon$ and $p$ by constants even in the case of the SB gas. These constant shifts are inevitable because we require the continuity of $p$ at the transition between high and low temperature phases, as implemented in Eq.(11). These shifts cause $O(1/T^4)$ deviations in $\epsilon/T^4$ and $p/T^4$ from the naive SB limits computed without constant terms. Because the entropy density $s$ is determined solely from effective degrees of freedom of the system irrespective of the vacuum structure etc., we do not have the freedom to shift $s$. This imposes that the shifts in $\epsilon$ and $p$ should cancel each other such that $\epsilon + p$ is left untouched. Hence, the relation $\epsilon = 3p$ for a SB gas is also violated through this shift, and tails of $\epsilon/T^4$, $p/T^4$ and $(\epsilon - 3p)/T^4$ in the high temperature phase do not reflect the width of the transition region directly. The width of the transition region will be tested through the behavior of quantities defined through the derivative of $s(T)$, such as the sound velocity [48] and the heat capacity $C_V = T \partial s/\partial T$.

Asakawa and Hatsuda also showed for massless SB gases both in high and low temperature phases that $\epsilon/T^4$ always overshoots the naive SB value near $T_c$. The behavior of $\epsilon/T^4$ shown in Fig. 13 for the SU(3) gauge theory therefore suggests that gluons acquire an effective mass [49].

### 4.3. Gluon propagator at high temperatures

Finite gluon mass at $T > 0$ is expected also from a perturbative calculation. The one-loop gluon self-energy for SU($N_C$) QCD produces a mass for the electric channel of the gluon propagator (the electric mass)

$$m_{el}(T) = \sqrt{(2N_C + N_F)/6} \, g(T) \, T \qquad (12)$$

which causes the Debye screening of color electric charges. It is expected that the magnetic channel of the gluon propagator also acquires a finite mass (the magnetic mass), although the one-loop perturbation theory fails to give a non-zero result: Collecting the leading IR singularities [50], or by a study of a high temperature effective theory [51], we expect

$$m_{mag}(T) \approx O(g^2(T) T). \qquad (13)$$

These masses are important to control the infrared behavior in calculations of some thermodynamic properties of the quark gluon plasma, such as transport coefficients. Because perturbation theory suffers from IR divergences at higher orders [52], a nonperturbative study is required to confirm these predictions.

In the low temperature phase, however, the confinement makes the shape of the gluon propagator non-trivial. Previous numerical studies of the SU(3) gauge theory in the Landau gauge [53–56] reported that effective gluon masses increase with time separation. This implies the appearance of negative norm states. Several fits were attempted, so far without definite conclusions.

This year, two studies in the high temperature phase were presented. Nakamura *et al.* simulated the SU(3) gauge theory at $\beta = 6.8$ on $48^3 \times 16$ and $48^3 \times 64$ lattices ($T/T_c \simeq 1.43$ and 0.48, respectively) using the stochastic gauge fixing [57], and Heller, Karsch and Rank studied the SU(2) gauge theory in the Landau gauge mainly on a $32^3 \times 8$ lattice at $\beta = 2.6 - 3.47$ ($T/T_c \simeq 1.3 - 16$) [58]. Both observed the increase of effective screening masses of gluons with spatial separation $z$ also in the high temperature phase. Nakamura *et al.* reported that the behavior of a gluon propagator is quite different at large $z$ between the low and the high temperature phases, presumably due to the lack of confinement in the latter phase. Heller *et al.* studied the electric and magnetic channels separately and computed $m_{el}$ and $m_{mag}$ in the high temperature phase.[2] For the magnetic mass, they found a good agreement with the expected $T$ dependence (13), $m_{mag}(T) = 0.466(15)\,g^2(T)\,T$, where the two-loop formula for $g(T)$ is used with $\Lambda = 0.262(18)\,T_c$ ($g^2(T) \simeq 4.1 - 2.2$ for $T/T_c = 2 - 16$). On the other hand, the electric mass, which is believed to be more reliably calculated [59], does not show the expected behavior (12) at all, even at $T \simeq 16T_c$. More work is required to clarify this puzzling phenomenon. Especially, a better understanding of the $z$ dependence of gluon propagators which includes a clearer identification of plateaus will be crucial.

### 4.4. Hadronic matter below $T_c$

From a quenched QCD simulation with staggered quarks on a $32^3 \times 8$ lattice, Boyd *et al.* reported that physical quantities such as $\langle \bar{\Psi}\Psi \rangle$, $m_\pi$, $f_\pi$, $m_{f_0}$, $m_\rho$ and probably $m_N$, do not show any significant $T$ dependence up to $T = 0.92T_c$ [60]. Because the allowed $T$ dependence from their data is much smaller than that expected previously from calculations using the chiral perturbation theory [61] and the Nambu-Jona-Lasinio model [62], this makes it more difficult to extract a phenomenologically clear evidence of a hot hadronic matter in heavy ion collisions. However, at the conference, Hioki discussed [63] that, even though pions etc. do not show a significant $T$ dependence, the topological charge may have an observable $T$ dependence, which should affect the $\eta'$ meson mass [64,65]. By quenched simulations at $\beta = 5.89$ on $16^3 \times 6$, $16^3 \times 8$ and $16^3 \times 16$ lattices ($T/T_c \simeq 0.93$, 0.75 and 0, respectively), he found evidence for a $T$ dependence of the topological charge distribution. This information can be used to compute $m_{\eta'}$. His results suggest that there exists an observable $T$ dependence in $m_{\eta'}$.

### 4.5. Valence quark chiral condensate

The Columbia group extended their previous study [66] of valence quark mass dependence in $N_F = 2$ QCD with staggered quarks using $16^3 \times 4$ and $32^3 \times 4$ lattices [67]. They studied the valence quark chiral condensate $\langle \bar{\zeta}\zeta \rangle$ as a function of the valence quark mass $m_{\rm val}$ and the sea quark mass $m_q$. $\langle \bar{\zeta}\zeta \rangle$ essentially measures the density of Dirac eigenstates with eigenvalues of $O(m_{\rm val})$. It can be shown that $\langle \bar{\zeta}\zeta \rangle(m_{\rm val} = 0, m_q) = 0$ is a sufficient condition for the restoration of chiral symmetry. Their data show that $m_{\rm val}$ dependence of $\langle \bar{\zeta}\zeta \rangle$ is actually sensitive to the phase. They also presented results for $\langle (\int dx \bar{\psi}\gamma_5\tau_i\psi)^2 \rangle - \langle (\int dx \bar{\psi}\tau_i\psi)^2 \rangle$ in the vicinity of $T_c$ which suggest that the axial $U_A(1)$ symmetry is broken just above $T_c$.

## 5. FINITE CHEMICAL POTENTIAL

In spite of phenomenological urgency, QCD at finite chemical potential $\mu$ on the lattice [68] remains to be a challenging issue because of the complex action problem [8]. The standard argument suggests that, at $T = 0$, the chiral symmetry is restored at $\mu_c \approx m_N/3$ where $m_N$ is the nucleon mass [69].

On the other hand, quenched QCD simulations, which ignore the fermion determinant containing the complex phase, show pathological behavior at $m_\pi/2 \lesssim \mu \lesssim m_N/3$ for $T = 0$ [69–71]: Finite baryon number and energy densities appear at $\mu \gtrsim m_\pi/2$, and $\langle \bar{\Psi}\Psi \rangle$, which is expected to be independent of $\mu$ below $\mu_c$, begins to decrease at $\mu \simeq m_\pi/2$. For $\mu \gtrsim m_N/3$, a recent study at $\beta = 0$ shows that all observables have their limiting vales: $\langle \bar{\Psi}\Psi \rangle = 0$ etc. [71]. Simulations for $m_\pi/2 \lesssim \mu \lesssim m_N/3$ require a large amount of the CPU time due to large fluctuations in observables and slow convergence of the algorithm used to invert fermion matrices. To understand this unexpected behavior in the interval $m_\pi/2 \lesssim \mu \lesssim m_N/3$, the effect of the quenching approximation [72], artifacts from finite lattices [73], and problems due to the use of staggered quarks [74] are examined, so far, however, without a definite conclusion.

---
[2] Because the effective masses are slowly increasing with $z$, they performed the fits by selecting data at large separations ($zT > 1$) assuming that plateaus are approximately reached there. Propagators as a function of the momentum squared $k^2$, which were reported to be statistically stable [55,56], were not studied.

In order to see if this unexpected behavior is caused by a physical transition at $\mu \approx m_\pi/2$, it is important to clarify the role of dynamical pions at finite chemical potential [70,71]. Recently, Hands et al. performed a full simulation of the 2+1 D Gross-Neveu model with staggered fermions at $\mu > 0$ [75]. Although the model lacks the feature of confinement, it contains a massless pion at $\mu < \mu_c$ in the chiral limit as a mesonic bound state of elementary fermions. An important point is that the fermion determinant of the theory is real and positive semidefinite also at $\mu > 0$. The results of their simulation are completely consistent with naive expectations and a $1/N_F$ calculation: no unexpected behavior is observed at $\mu \simeq m_\pi/2$. This suggests that the pathological behavior of quenched QCD at $m_\pi/2 \lesssim \mu \lesssim m_N/3$ is not physical. Although other possibilities are not excluded [73,74], it is natural to expect that a full QCD simulation will lead us to a simple phase structure. At present, full QCD simulations are attempted only on small lattices (typically $4^4$) and/or in the strong coupling limit [76–80]. More efforts are needed to solve the problem of complex actions.[3]

## 6. CONCLUSIONS AND PROSPECTS

I reviewed recent progress in finite temperature lattice QCD. We found evidences that the scaling properties near the chiral transition in two flavor QCD are consistent with the O(4) scaling both with staggered and Wilson quarks. It was pointed out that use of some improved action is essential to obtain the scaling behavior for Wilson quarks on small $N_t$ lattices (presumably for $N_t \lesssim 18$), by eliminating lattice artifacts with the standard action. Also a study on the structure of the critical line at $T > 0$ with Wilson quarks was reported. These progresses provide us with more rigid conceptual and technical bases for Wilson quarks. For more confirmation of the O(4) scaling, further studies and better accuracy of data are required: For staggered quarks, data on larger lattices and at smaller quark masses are needed to check consistency with the O(4) scaling, and, in future, to distinguish between O(4) and O(2) exponents. For Wilson quarks, a direct determination of critical exponents should be attempted.

---
[3] A different approach, the static approximation, was presented at the conference [81].

I also reviewed the status of investigations in QCD including the strange quark. Although both staggered and Wilson simulations performed on $N_t = 4$ lattices give a phase structure qualitatively consistent with the expected one shown in Fig. 7, Wilson quarks tend to give larger critical quark masses in comparison with staggered quarks. This consequently leads to the difference in the conclusion about the location of the real world in Fig. 7. Because the deviation from the continuum limit is large in the both simulations at $N_t = 4$, we should certainly make a calculation at larger $N_t$ or with an improved action in order to draw a definite conclusion about the nature of the transition in the real world.

I finally reviewed the progress in other topics of finite temperature/density QCD. The results are both encouraging and challenging and it is important to perform further studies in these directions. Improvement of the lattice action may prove to be useful also in these studies.

I am indebted to S. Aoki, T. Blum, S. Chandrasekharan, N. Christ, T. Hatsuda, S. Hioki, F. Karsch, A. Kocić, J.B. Kogut, E. Laermann A. Nakamura, D.K. Sinclair, and A. Ukawa for valuable discussions and providing me with original data. I am also grateful to my collaborators, Y. Iwasaki, S. Kaya, S. Sakai, and T. Yoshié for their support and encouragements. Finally I would like to thank Y. Iwasaki, A. Ukawa and W. Bock for critical comments and suggestions on the manuscript. This work is in part supported by the Grant-in-Aid of Ministry of Education, Science and Culture (Nos.07NP0401 and 07640376).

## REFERENCES

1. A. Ukawa, Nucl. Phys. B (Proc. Suppl.) 17 (1990) 118.
2. F. Karsch, Nucl. Phys. B (Proc. Suppl.) 34 (1994) 63.

3. R. Pisarski and F. Wilczek, Phys. Rev. D29(1984) 338.
4. F. Wilczek, Int. J. Mod. Phys. A7(1992) 3911; K. Rajagopal and F. Wilczek, Nucl. Phys. B399(1993) 395.
5. Y. Iwasaki, Nucl. Phys. B (Proc. Suppl.) 42 (1995) 96.
6. F.R. Brown, et al., Phys. Rev. Lett. 65(1990) 2491.
7. K. Rajagopal, in *Quark-Gluon Plasma 2*, ed. R. Hwa, World Scientific, 1995.
8. I.M. Barbour, Nucl Phys. B (Proc. Suppl.) 26 (1992) 22.
9. C. DeTar, Nucl. Phys. B (Proc. Suppl.) 42 (1995) 73; in *Quark-Gluon Plasma 2*, ed. R. Hwa, World Scientific, 1995.
10. K. Kanaya and S. Kaya, Phys. Rev. D51 (1995) 2404; A recent study with high temperature series to the 19th order by P. Buetera and M. Comi also gives completely consistent exponents (private communication).
11. J.C. Le Guillou and J. Zinn-Justin, Phys. Rev. B21 (1980) 3976; J. Phys. Lett. (Paris) 46 (1985) L-137.
12. F. Karsch, Phys. Rev. D49 (1994) 3791; F. Karsch and E. Laermann, *ibid.* D50 (1994) 6954.
13. S. Gottlieb, et al., Phys. Rev. D35 (1987) 3972; M. Fukugita, H. Mino, M. Okawa and A. Ukawa, Phys. Rev. Lett. 65 (1990) 816; Phys. Rev. D42 (1990) 2936; C. Bernard, et al., Phys. Rev. D45 (1992) 3854.
14. M. Fukugita, H. Mino, M. Okawa and A. Ukawa, Phys. Rev. Lett. 65 (1990) 816; Phys. Rev. D42 (1990) 2936.
15. A. Vaccarino, Nucl. Phys. B (Proc. Suppl.) 20 (1991) 263.
16. E. Laermann, these proceedings.
17. G. Boyd, et al., Nucl. Phys. B376 (1992) 199.
18. A. Kocić and J. Kogut, Phys. Rev. Lett. 74 (1995) 3112; preprint hep-lat/9507012.
19. M. Bochicchio, et al., Nucl. Phys. B262(1985) 331.
20. S. Itoh, Y. Iwasaki, Y. Oyanagi and T. Yoshié, Nucl. Phys. B274(1986) 33.
21. Y. Iwasaki, K. Kanaya, S. Sakai and T. Yoshié, Phys. Rev. Lett. 67 (1991) 1494; Y. Iwasaki, T. Tsuboi and T. Yoshié, Phys. Lett. B220 (1989) 602.
22. C. Bernard, et al., Phys. Rev. D49 (1994) 3574.
23. Y. Iwasaki, K. Kanaya, S. Sakai and T. Yoshié, Tsukuba preprint UTHEP-300(1995); Nucl. Phys. B (Proc. Suppl.) 30 (1993) 327; *ibid.* 34 (1994) 314; *ibid.* 42 (1995) 499.
24. Y. Iwasaki, K. Kanaya, S. Sakai and T. Yoshié, Phys. Rev. Lett. 69 (1992) 21; S. Itoh, Y. Iwasaki and T. Yoshié, Phys. Rev. D36 (1986) 527.
25. S. Aoki, A. Ukawa and T. Umemura, Tsukuba preprint UTHEP-313 (1995); in these proceedings.
26. S. Aoki, Phys. Rev. D30 (1984) 2653; Phys. Rev. Lett. 57 (1986) 3136; Nucl. Phys. B314 (1989) 79; Tsukuba preprint UTHEP-318 (1995).
27. C. Bernard, et al., Phys. Rev. D46 (1992) 4741; T. Blum, et al., *ibid.* D50 (1994) 3377.
28. Y. Iwasaki, K. Kanaya, S. Sakai, and T. Yoshié, Nucl. Phys. B (Proc. Suppl.) 42 (1995) 502; K. Kanaya, Tsukuba preprint UTHEP-306 (1995).
29. T. Yoshié, these proceedings.
30. Y. Iwasaki, Nucl. Phys. B258 (1985) 141; Tsukuba preprint UTHEP-118 (1983) unpublished.
31. L. Maiani and G. Martinelli, Phys. Lett. B178 (1986) 265; D. Daniel, et al., Phys. Rev. D46 (1992) 3130.
32. A. Hoferichter, V.K. Mitrjushkin and M. Müller-Preussker, preprint HU Berlin-IEP-95/5.
33. R.V. Gavai and F. Karsch, Nucl. Phys. B261(1985) 273; R.V. Gavai, J. Potvin and S. Sanielevici, Phys. Rev. Lett. 58(1987) 2519.
34. J.B. Kogut and D.K. Sinclair, Phys. Rev. Lett. 60 (1988) 1250; Phys. Lett. B229 (1989) 107; Nucl. Phys. B344 (1990) 238.
35. Y. Iwasaki, K. Kanaya, S. Kaya, S. Sakai and T. Yoshié, Tsukuba preprint UTHEP-304(1995); Nucl. Phys. B (Proc. Suppl.) 42 (1995) 499.
36. J.B. Kogut, D.K. Sinclair and K.C. Wang, Phys. Lett. B263 (1991) 101.
37. G. Boyd, et al., Bielefeld preprint BI-TP

95/23 (1995); E. Laermann, these proceedings.
38. S.A. Gottlieb, et al., Phys. Rev. D46 (1992) 4657; N.H. Christ and A.E. Terrano, Phys. Rev. Lett. 96 (1986) 111; A.D. Kennedy, J. Kuti, S. Meyer and B.J. Pendleton, Phys. Rev. Lett. 54 (1985) 87.
39. J. Fingberg, U. Heller and F. Karsch, Nucl. Phys. B392 (1993) 493.
40. M. Fukugita, M. Okawa and A. Ukawa, Nucl. Phys. B337 (1990) 181.
41. Y. Iwasaki, et al., Phys. Rev. D46 (1992) 4657; Phys. Rev. Lett. 67 (1991) 3343.
42. T. Blum, these proceedings.
43. T. Blum, L. Kärkkäinen, D. Toussaint and S. Gottlieb, Phys. Rev. D51 (1995) 5153.
44. C. DeTar, these proceedings.
45. J. Engels, et al., Phys. Lett. B525 (1990) 625.
46. K. Akemi, et al., Phys. Rev. Lett. 71 (1993) 3063.
47. M. Asakawa and T. Hatsuda, Columbia preprint CU-TP-705 (1995).
48. L. Kärkkäinenn, these proceedings.
49. T. Hatsuda, private communication.
50. A.D. Linde, Rep. Prog. Phys. 42 (1979) 389.
51. D. Gross, R.D. Pisarski and L.G. Yaffe, Rev. Mod. Phys. 53 (1981) 43.
52. A.D. Linde, Phys. Lett. B96 (1980) 289.
53. J.E. Mandula and M. Ogilvie, Phys. Lett. B185 (1987) 127.
54. R. Gupta, et al., Phys. Rev. D36 (1987) 2813.
55. C. Bernard, C. Parrinello and A. Soni, Phys. Rev. D49 (1994) 1585.
56. P. Marenzoni, G. Martinelli, N. Stella and M. Testa, Phys. Lett. B318 (1993) 511; P. Marenzoni, G. Martinelli and N. Stella, Rome preprint ROME prep. 94/1042.
57. A. Nakamura, et al., Yamagata preprint YAMAGATA-HEP-95-10.
58. U.M. Heller, F. Karsch and J. Rank, Bielefeld preprint BI-TP 95/21 (1995).
59. See, for example, J.-P. Blaizot, J.-Y. Ollitrault and E. Iancu, in Quark-Gluon Plasma 2, ed. R. Hwa, World Scientific, 1995.
60. G. Boyd, et al., Nucl. Phys. B (Proc. Suppl.) 42 (1995) 469; Phys. Lett. B349 (1995) 170.
61. P. Gerber and H. Leutwyler, Nucl. Phys. B321 (1989) 387.
62. T. Hatsuda and T. Kunihiro, Phys. Rev. Lett. 55 (1985) 158; Phys. Lett. B185 (1987) 304; ibid. B198 (1987) 126.
63. S. Hioki, these proceedings.
64. S. Itoh, Y. Iwasaki and T. Yoshié, Phys. Rev. D36 (1987) 527.
65. M. Fukugita, Y. Kuramashi, M. Okawa and A. Ukawa, Phys. Rev. D51 (1995) 3952.
66. S. Chandrasekharan, Nucl. Phys. B (Proc. Suppl.) 42 (1995) 475.
67. S. Chandrasekharan and N. Christ, these proceedings.
68. J. Kogut, et al., Nucl. Phys. B225 (1983) 93; P. Hasenfratz and F. Karsch, Phys. Lett. B125 (1983) 308.
69. I. Barbour, et al., Nucl. Phys. B275 [FS17] (1986) 296.
70. C.T.H. Davies and E.G. Klepfish, Phys. Lett. B256 (1991) 68.
71. J.B. Kogut, M.-P. Lombardo and D.K. Sinclair, Phys. Rev. D51 (1995) 1282; Nucl. Phys. B (Proc. Suppl.) 42 (1995) 514.
72. P.E. Gibbs, Phys. Lett. B182 (1986) 369; A. Gocksch, Phys. Rev. D37 (1988) 1014. See, however, a critical note: N. Bilić and K. Demeterfi, Phys. Lett. B212 (1988) 83.
73. E. Mendel, Nucl. Phys. B (Proc. Suppl.) 4 (1988) 308; N. Bilić and K. Demeterfi, in [72]; J.C. Vink, Nucl. Phys. B323 (1989) 399.
74. E. Mendel, Nucl. Phys. B387 (1992) 485. For recent status, see E. Mendel and L. Polley, ibid. B (Proc. Suppl.) 42 (1995) 535; W. Wilcox, S. Trendafilov and E. Mendel, ibid. B (Proc. Suppl.) 42 (1995) 557.
75. S. Hands, S. Kim and J.B. Kogut, Nucl. Phys. B442 (1995) 364.
76. I.M. Barbour, C.T.H. Davies and Z. Sabeur, Phys. Lett. B215 (1988) 567; I.M. Barbour and Z.A. Sabeur, Nucl. Phys. B342 (1990) 269; A. Hasenfratz and D. Toussaint, ibid. B371 (1992) 539.
77. A. Gocksch, Phys. Rev. Lett. 61 (1988) 2054.
78. D. Toussaint, Nucl. Phys. B (Proc. Suppl.) 17 (1990) 248.
79. F. Karsch and H.W. Wyld, Phys. Rev. Lett. 55 (1985) 2242; N. Bilić, H. Gausterer and S. Sanielevici, Phys. Lett. B198 (1987) 235.
80. E. Dagotto, A. Moreo and U. Wolff, Phys.

Rev. Lett. 57 (1986) 1292; F. Karsch and K.-H. Mütter, Nucl. Phys. B313 (1989) 541.
81. T.C. Blum, J.E. Hetric and D. Toussaint, Arizona preprint AZPH-TH/95-22; these proceedings.